\documentclass{article}

\usepackage{graphicx}   % For \includegraphics
\usepackage{subcaption} % For the subfigure environment and \captionsetup
\usepackage{pgffor}     % For the \foreach loop
\usepackage{arxiv}
\usepackage{orcidlink}
\usepackage{pdflscape}
\usepackage[utf8]{inputenc} % allow utf-8 input
\usepackage[T1]{fontenc}    % use 8-bit T1 fonts
\usepackage{hyperref}       % hyperlinks
\usepackage{url}            % simple URL typesetting
\usepackage{booktabs} 
\usepackage{longtable}% professional-quality tables
\usepackage{amsfonts}       % blackboard math symbols
\usepackage{nicefrac}       % compact symbols for 1/2, etc.
\usepackage{microtype}      % microtypography
\usepackage{lipsum}
\usepackage{graphicx}
\usepackage{amsmath}
\usepackage{adjustbox}
\usepackage{xcolor}
\usepackage{multirow}
\graphicspath{ {./images/} }

\title{Breaking Rank - A Novel Unscented Kalman Filter for Parameter Estimations of a Lumped-parameter Cardiovascular Model}

\author{
Alex Thornton\\
    School of Computer Science\\
    Insigneo Institute for in silico Medicine\\
    University of Sheffield \\
    Sheffield, United Kingdom \\
   \And
 Ian Halliday \\
    Clinical Medicine, The Medical School\\
    NIHR Sheffield Biomedical Research Centre\\
    Insigneo Institute for in silico Medicine\\
    University of Sheffield \\
    Sheffield, United Kingdom \\
   \And  
 Harry Saxton \orcidlink{0000-0001-7433-6154}\\
    Astellas Pharma\\
    United Kingdom\\
   \And
 Xu Xu \orcidlink{0000-0002-9721-9054}\\
    School of Computer Science\\
    Insigneo Institute for in silico Medicine\\
    University of Sheffield \\
    Sheffield, United Kingdom \\
    xu.xu@sheffield.ac.uk
}

\begin{document}
\maketitle
\begin{abstract}
 We make modifications to the unscented Kalman filter (UKF) which bestow almost complete practical identifiability upon a lumped-parameter cardiovascular model with 10 parameters and 4 output observables - a highly non-linear, stiff problem of clinical significance. The modifications overcome the challenging problems of rank deficiency when applying the UKF to parameter estimation. Rank deficiency usually means only a small subset of parameters can be estimated. Traditionally, pragmatic compromises are made, such as selecting an optimal subset of parameters for estimation and fixing non-influential parameters. Kalman filters are typically used for dynamical state tracking, to facilitate the control $u$ at every time step. However, for the purpose of parameter estimation, this constraint no longer applies. Our modification has transformed the utility of UKF for the parameter estimation purpose, including minimally influential parameters, with excellent robustness (i.e., under severe noise corruption, challenging patho-physiology, and no prior knowledge of parameter distributions). The modified UKF algorithm is robust in recovering almost all parameters to over 98\% accuracy, over 90\% of the time, with a challenging target data set of 50, 10-parameter samples. We compare this to the original implementation of the UKF algorithm for parameter estimation and demonstrate a significant improvement.

\end{abstract}

% keywords can be removed

\section{Introduction}
\label{sec:intro}
Zero-dimensional (0D) lumped-parameter (LP) cardiovascular models have been used for over a hundred years \cite{frank1899grundform}. They eliminate the spatial domain of mathematical modelling, which vastly reduces the computational complexity of simulations. This allows millions of simulations to be run (for example, for global sensitivity analysis), where in higher spatial dimensions it would only be possible to run a few \cite{saxton2025uncertainty}. Statistical Models are widely used in healthcare settings. Factors such as smoking, family history of illness and age, change the conditional probability of having an underlying pathophysiology, which is very useful information for clinicians. The advantage of using mechanistic models over statistical models is that these models produce predictions based on physical principles and biological processes; hence, they can be precisely tuned to represent realistic physiology and provide deeper insight. Much of the mechanistic modelling community's effort in recent years has been on how best to calibrate these models, so they can be personalised to represent each patient's specific physiological conditions. There is a family of prominent engineering and computational techniques for this purpose, from gradient descent, Markov chain Monte Carlo, to Kalman filter-like methods \cite{arnold2018approach,qureshi2019hemodynamic}.

The Kalman filter family contains a broad range of algorithms designed for different systems. Originally developed for optimal state estimation in the context of guidance, navigation and control, within stochastic environments, typically characterised by measurement noise and process noise. The fundamental Kalman filter (KF) provides an optimal linear recursive solution for estimating the state of linear dynamical systems perturbed by white Gaussian noise \cite{Kalman_1960,bishop2001introduction}. Modifications to this algorithm have been developed over the years to better deal with nonlinear system dynamics. For example, the extended Kalman filter (EKF) uses a first-order Taylor series linearisation around the current state estimate, which improves performance for systems with mild non-linearities \cite{bishop2001introduction}. For high-dimensional, complex nonlinear systems like weather forecasts, the ensemble Kalman filter (EnKF) was developed. The EnKF propagates a set of state estimates (an ensemble) through the nonlinear dynamics and estimates the required covariance matrix directly from this ensemble, and is also broadly used in cardiovascular modeling and other areas of data assimilation \cite{burgers1998analysis, arnold2014parameter, knapp2025personalizing}. The unscented Kalman filter (UKF) is a more recent addition to the KF family. The UKF employs the unscented transform to deterministically select a minimal set of sample points (sigma points) whose propagation through the true nonlinear system captures the posterior mean and covariance accurately to at least the second order, which results in a more robust estimate than the EKF. This makes it naturally suitable for complex problems such as cardiovascular modeling \cite{wan2000unscented,muller2018reduced}.

Applications of the UKF to parameter estimation to date align with the originally intended purpose of the KF family: simultaneous parameter and state tracking as new observations become available at each time step. Marchesseau et. al utilised the UKF to estimate four global haemodynamic parameters: myocardial contractility, blood viscosity, aortic bulk modulus and peripheral resistance within an aortic model, fixing the remaining ten parameters to nominal physiological values \cite{marchesseau2013personalization}. More recently, Pant et. al \cite{pant2014methodological, pant2017inverse} employed the UKF to estimate Windkessel model parameters, but found the results were highly sensitive to initial values chosen for the prior state estimate $\mathbf{\hat{x}_{0|0}}$ and the initial error covariance matrix $\mathbf{P_{0|0}}$. Saxton et. al tried to recover 9-parameters under the influence of a varying heart rate with the UKF, with initial guesses at 10\% away from known values, and found that 6 of the parameters were recoverable. 

Caiazzo et. al, used a reduced order UKF to identify parameters related to arterial wall stiffness and thickness within a one-dimensional (1D) blood flow model. The authors implemented the method using only one experimental dataset and noted that the UKF exhibited divergence from the established reference values \cite{caiazzo2017assessment}. Subsequently, Muller and Caiazzo et. al investigated the use of frequency domain measurements to improve the performance of the filter, which showed promising results in one example \cite{muller2018reduced}.  

There are many issues that researchers have encountered when attempting to estimate parameters, in even the simplest models, using clinical or quasi-clinical data. This limitation in model identifiability currently limits the clinical utility of LP models. It is widely accepted that even the simplest models suffer from rank deficiency \cite{canuto2020ensemble}. Rank deficiency is best described as a situation where the observed data, which are used to fit the model against target clinical data, lack sufficient information content to uniquely determine all model parameters. This results in the non-uniqueness of the inverse solution (non-identifiability) \cite{colebank2024guidelines}. Rank deficiency is a prevalent challenge across many fields, often interpreted as reflecting fundamental limits to the achievable level of model identifiability. This leads to compromises, such as fixing the least sensitive parameters to nominal physiological values, which can be problematic when personalising models and building patient-specific digital twins \cite{yu2022overall,canuto2020ensemble}. 

% The following sections of the paper will contain the explanations of the modifications to the UKF for the purpose of parameter estimation. The paper will present results demonstrating the generalisability and robustness of this method in recovering parameters of the 10-parameter cardiovascular model with clinical considerations in mind. The results will also demonstrate the method's ability to infer blind states. Then the clinical relevance will then be discussed and the results presented in full laterly in the appendix. 

We address these issues through modifications of the original UKF in this paper. We apply this modified UKF to a 10-parameter 0D cardiovascular circulation model and present an ensemble of parameter estimation results to demonstrate the generalisability and robustness of the new method. 

The structure of the paper is as follows. In section 2, we provide an overview of the original UKF for parameter estimation. We then explain the modifications to the UKF method that are possible when the primary objective is the estimation of parameters as opposed to the dynamic state tracking for the purpose of control. In section 3, we present the application of the modified UKF formulation to estimate parameters of a 10-parameter LP cardiovascular model. In section 4, we present an ensemble of estimation results which demonstrate the ability of the new method to overcome issues related to rank deficiency and estimate all ten parameters with a high degree of accuracy and reliability. We evaluate the robustness of the modified UKF across several critical factors, including resilience to measurement noise, its performance under conditions of unknown prior parameter distributions, its generalisability and overall reliability. Finally, we present comparisons with the original implementation of the UKF algorithm for parameter estimation, thereby demonstrating the significant advancement achieved in this field. 

\section{Methods}

\subsection{The Original UKF for Parameter Estimation}
\label{sec: original_ukf}

A general discrete-time dynamical system at time $k-1$ is governed by:
\begin{equation}
\label{eq:state}
     \mathbf{x}_{state,k} = f_{state}(\mathbf{x}_{state, k-1}, \mathbf{\theta}_{k-1}, \mathbf{u}_{k-1}, \mathbf{w}_{state,k-1}),
\end{equation}
with the observation equation :
\begin{equation}
    \label{eq:observation}
    \mathbf{y}_{k-1} = h_{obs}(\mathbf{x}_{state,k-1}, \mathbf{v}_{k-1}) ,
\end{equation}
where $\mathbf{x}_{state} \in \mathbb{R}^m $, $\mathbf{\theta} \in \mathbb{R}^p$, $\mathbf{u} \in \mathbb{R}^m$ and $\mathbf{y} \in \mathbb{R}^n$ denote the internal state vector, the parameter vector, the control input vector and the observed output vector, respectively. $f_{state}$ and $h_{obs}$ are the state transition and measurement functions, and $\mathbf{w}_{state}$ and $\mathbf{v}$ represent the process and measurement noises.

Here, the goal is to estimate the full set of parameters, $\theta$, within the nonlinear system defined in Eq.~(\ref{eq:state}), from the measured outputs $\mathbf{y}$. Below, we summarise the key steps of state and parameter estimations using UKF, following the seminal works of Julier and Uhlmann \cite{julier2004unscented}, Wan and van der Merwe \cite{wan2000unscented} and Pant et. al \cite{pant2014methodological}. As UKF is not used to facilitate control here, we neglect the term $\mathbf{u}$ in the rest of the paper. 

\subsubsection{System Augmentation}
{\bf In this section, all symbols have their usual meaning, unless explictly stated otherwise.} To estimate parameters, we create an augmented state vector, $\mathbf{x}$, which includes both the internal states and the parameters:
$$
\mathbf{x}_k = 
\begin{bmatrix} 
\mathbf{x}_{state, k} \\ 
\mathbf{\theta}_k 
\end{bmatrix},
$$
where $\mathbf{x} \in \mathbb{R}^L$, (i.e., $L = m+p$). The system can now be re-defined as:
\begin{align*}
    \mathbf{x}_k &= f(\mathbf{x}_{k-1} , \mathbf{w}_{k-1}) = 
    \begin{bmatrix} 
        f_{state}(\mathbf{x}_{state, k-1}, \mathbf{\theta}_{k-1},  \mathbf{w}_{state, k-1}) \\ 
        \mathbf{\theta}_{k-1} + \mathbf{\eta}_{k-1} 
    \end{bmatrix} \\
    \mathbf{y}_{k-1} &= h(\mathbf{x}_{k-1}, \mathbf{v}_{k-1})  = h_{obs}(\mathbf{x}_{state, k-1}, \mathbf{v}_{k-1}  )   
\end{align*}
where $\mathbf{\eta}_{k-1}$ represents a small amount of noise for the parameters which allows $\theta$ to be updated.

\subsubsection{Initialisation, Sigma Point Generation \& Weights}
We initialise the augmented prior state estimate $\mathbf{\hat{x}}_{0|0}$ and its error covariance matrix $\mathbf{P}_{0|0}$ as follows:
$$
\mathbf{\hat{x}}_{0|0} = 
\begin{bmatrix} 
    \mathbf{\hat{x}}_{state, 0} \\ 
    \mathbf{\hat{\theta}}_0 
\end{bmatrix}, \quad
\mathbf{P}_{0|0}  = 
\begin{bmatrix} 
    \mathbf{P}_{state, 0} & 0 \\ 
    0 & \mathbf{P}_{\theta, 0}. 
\end{bmatrix}
$$
Then we generate $2L+1$ sigma points $\mathbf{\mathcal{X}}_i$ with corresponding weights $W_i$ as follows:
$$
\mathbf{\mathcal{X}}_{0, k-1|k-1} = \mathbf{\hat{x}}_{k-1|k-1}
$$
$$
\mathbf{\mathcal{X}}_{i, k-1|k-1} = \mathbf{\hat{x}}_{k-1|k-1} \pm \left( \sqrt{(L+\lambda)\mathbf{P_{\tilde{x}\tilde{x}}}_{k-1|k-1}} \right)_i, \quad \text{for \ } i = 1, \dots, L
$$
where $\left( \sqrt{A} \right)_i$ is the $i$-th column of the matrix square root of $A$ (e.g., using Cholesky decomposition as a numerical approximation in simulations).
The scaling parameter $\lambda$ is defined as $\lambda = \alpha^2(L+\kappa) - L$, where $\alpha$ controls the spread of the sigma points (e.g., $10^{-3}$) and $\kappa$ is a secondary scaling parameter to give further flexibility for the spread of sigma points. The weights for the mean ($W^{(m)}$) and covariance ($W^{(c)}$) are calculated as
$ W_0^{(m)} = \frac{\lambda}{L+\lambda}$, $W_0^{(c)} = \frac{\lambda}{L+\lambda} + (1 - \alpha^2 + \beta)$ and $W_i^{(m)} = W_i^{(c)} = \frac{1}{2(L+\lambda)}$ for $i = 1, \dots, 2L$, where $\beta$ incorporates prior knowledge of the distribution ($\beta=2$ is optimal for Gaussian distribution).

\subsubsection{Prediction (\textit{a priori} Estimate)}
Next, the sigma points are propagated through the augmented state function $f$ to calculate the predicted augmented state mean $\mathbf{\hat{x}}$ and error covariance $\mathbf{P_{\tilde{x}\tilde{x}}}$ at step $k$, given information up to step $k-1$: 
\begin{align*}
    \mathbf{\mathcal{X}}_{i, k|k-1} &= f(\mathbf{\mathcal{X}}_{i, k-1|k-1},  \mathbf{w}_{k-1}), \quad \text{for \ } i = 0, \dots, 2L    \\
    \mathbf{\hat{x}}_{k|k-1} &= \sum_{i=0}^{2L} W_i^{(m)} \mathbf{\mathcal{X}}_{i, k|k-1}, \\
    \mathbf{P_{\tilde{x}\tilde{x}}}_{k|k-1} &= \sum_{i=0}^{2L} W_i^{(c)} \left( \mathbf{\mathcal{X}}_{i, k|k-1} - \mathbf{\hat{x}}_{k|k-1} \right) \left( \mathbf{\mathcal{X}}_{i, k|k-1} - \mathbf{\hat{x}}_{k|k-1} \right)^T. 
\end{align*}

\subsubsection{Correction (\textit{a posteriori} Estimate) }
In this stage, the predicted states from the previous stage are corrected. This is first performed through a preparation step of propagating the sigma points and calculating: (i) the predicted measurement mean vector $ \mathbf{\hat{y}}_{k|k-1}$, (ii) the innovation covariance matrix $\mathbf{P_{\tilde{y}\tilde{y}}}$, (iii) the cross-covariance matrix $\mathbf{P_{\tilde{x}\tilde{y}}}$ and (iv) the optimal Kalman gain matrix ($\mathbf{K}$):
\begin{align*}
    \mathbf{\mathcal{Y}}_{i, k|k-1} &= h(\mathbf{\mathcal{X}}_{i, k|k-1}, \mathbf{v}_{k-1}), \quad \text{for } i = 0, \dots, 2L, \\
    \mathbf{\hat{y}}_{k|k-1} &= \sum_{i=0}^{2L} W_i^{(m)} \mathbf{\mathcal{Y}}_{i, k|k-1}, \\
    \mathbf{P_{\tilde{y}\tilde{y}}}_{k|k-1} &= \sum_{i=0}^{2L} W_i^{(c)} \left( \mathbf{\mathcal{Y}}_{i, k|k-1} - \mathbf{\hat{y}}_{k|k-1} \right) \left( \mathbf{\mathcal{Y}}_{i, k|k-1} - \mathbf{\hat{y}}_{k|k-1} \right)^T + \mathbf{R}_k, \\
    \mathbf{P_{\tilde{x}\tilde{y}}}_{k|k-1} &= \sum_{i=0}^{2L} W_i^{(c)} \left( \mathbf{\mathcal{X}}_{i, k|k-1} - \mathbf{\hat{x}}_{k|k-1} \right) \left( \mathbf{\mathcal{Y}}_{i, k|k-1} - \mathbf{\hat{y}}_{k|k-1} \right)^T, \\
    \mathbf{K}_k &= \mathbf{P_{\tilde{x}\tilde{y}}}_{k|k-1} \mathbf{P_{\tilde{y}\tilde{y}}}_{k|k-1}^{-1},
\end{align*}
where $\mathbf{R}$ is the covariance matrix of the measurement noise $\mathbf{v}$.
 Then, the state mean is updated using the optimal Kalman Gain and the latest measurements $\mathbf{y}$ obtained at step $k$:
 \begin{equation*}
     \mathbf{\hat{x}}_{k|k} = \mathbf{\hat{x}}_{k|k-1} + \mathbf{K}_k (\mathbf{y}_k - \mathbf{\hat{y}}_{k|k-1}).
 \end{equation*}
Finally, we calculate the updated state error covariance, ready for the iteration at the next time step $k+1$:
 \begin{equation*}
     \mathbf{P_{\tilde{x}\tilde{x}}}_{k|k} = \mathbf{P_{\tilde{x}\tilde{x}}}_{k|k-1} - \mathbf{K}_k \mathbf{P_{\tilde{y}\tilde{y}}}_{k|k-1} \mathbf{K}_k^T.
 \end{equation*}

 The UKF uses a deterministic sampling approach whereby a set of sampling points (the sigma points) is chosen to capture the mean and covariance of the state probability distribution. These sigma points are propagated through the nonlinear system model, and then the summary statistics related to the prior error covariance of the state, the cross-covariance and the predicted measurement covariance are generated in matrix forms \cite{saxton2023personalised,wan2000unscented}. These summary statistics are then used to update the prediction of the mean and covariance to provide the posterior mean and covariance at the current estimation stage. The complete process, involves setting the sigma points, and then propagating them through the model. The information from these sigma points is used to calculate the cross-covariance matrix $\mathbf{P_{\tilde{x}\tilde{y}_{k|k-1}}}$ and the predicted measurement covariance matrix $\mathbf{P_{\tilde{y}\tilde{y}_{k|k-1}}}$, this is used to formulate a matrix called the Kalman gain. The Kalman gain is used to project forward the models state $\mathbf{\hat{x}}$ and covariance $\mathbf{P_{\tilde{x}\tilde{x}_{k|k}}}$ to refine it's estimate, this is performed recursively. This recursive algorithm was initially designed to better track objects in challenging conditions where the system dynamics were sensitive to initial conditions and only noisy observation data were available \cite{qi2016dynamic} but has recently found utility in parameter estimation \cite{saxton2023personalised} as summarised in this section.

\subsection{Implementation Challenges of the Original UKF}

As explained in Section \ref{sec: original_ukf}, the application of the UKF method to estimate system parameters involves the recursive updating of the augmented state vector $\mathbf{x} = [\mathbf{x}_s^{\mathsf{T}}, \mathbf{\theta}^{\mathsf{T}}]^{\mathsf{T}}$ at each measurement time instant. The filter attempts to converge upon the parameter set $\mathbf{\theta}$ that minimises the estimation error between the predicted and the observed system outputs. This methodology often encounters three primary challenges: 
\begin{enumerate}
    \item Observability and rank deficiency: Researchers have shown that the parameter estimation problem is fundamentally an underdetermined inverse problem which affects the uniqueness of the obtained solutions\cite{canuto2020ensemble,yeh1986review}. The filter often operates in a regime where the dimension of the measurement vector, $\mathbf{y}$, is significantly smaller than the dimension of the augmented state vector $\mathbf{x}$ \cite{khalili2024state,pant2017inverse}. Therefore, the nonlinear measurement function, $\mathbf{h}$, fails to provide sufficient information to distinguish changes in all state and parameter variables. This condition manifests as poor system observability, which can lead to the error covariance matrices becoming rank deficient (singular or nearly singular), hindering the necessary inversion step during the Kalman gain computation and the state estimate correction \cite{canuto2020ensemble}.
    
    \item Transient dynamics versus steady-state behaviour: The recursive prediction and update structure, particularly in models highly sensitive to initial conditions, inherently emphasises tracking the instantaneous system dynamics (the transition from $\mathbf{x}_{k} \to \mathbf{x}_{k+1}$) based on noisy observations $\mathbf{y}_{k}$. This focus on immediate transient behaviour can result in parameter estimates that are biased towards fitting short-term fluctuations and may fail to accurately reproduce the long-term (steady-state) dynamics of the system over extended prediction horizons.
    \item Performance highly dependent on particular initial conditions: A non-optimal choice of $\mathbf{\hat{x}_{0|0}}$ and $\mathbf{P_{0|0}}$ can lead to the failure of the UKF, even for less challenging dynamical systems. Therefore prior knowledge of parameter and initial state distributions has been necessary for the success of this method \cite{pant2014methodological,majeed2013aerodynamic}. 
\end{enumerate}

\subsection{The Modified UKF Approach}
\label{sec:modified_approach}

In this study, the UKF is applied to estimate parameters of a non-linear system, governed by a set of ODEs. This differs from the traditional utility of the UKF, where it was designed as a non-linear filter to assimilate new measurement information to estimate internal states of a system. Neither the observations nor the prior state predictions could be relied upon solely, because the observations were noisy and the state prediction were inaccurate due to model and process mismatch. The original UKF or other KF filters were employed to facilitate the control by providing the best current estimate of the state vector $\mathbf{x}$ for feedback. 

In the present application, however, we consider only deterministic systems (neglecting numerical errors) whose entire trajectory is uniquely determined by the set of initial state conditions and model parameters. However, these parameters, and in some cases, the initial state conditions are entirely unknowns. Furthermore, these systems do not require a nonlinear control input $\mathbf{u}$ which would necessitate the real-time modification of the dynamical system at every observation step to obtain the best estimate $\mathbf{x}$ for the purpose of applying $\mathbf{u}$. Under these constraints, the UKF method can be adapted as follows: 

% The old construction of the problem with the new implementation was problematic. This is due to the construction of the problem being what change in parameters at $y(t)$ best describe the change from $y(t) \xrightarrow{}y(t+dt)$ where $\delta t$ is the next observation point. This is not the solution to the problem given that both $y(t)$ and $y(t+dt)$ are noisy observations of the target solution, such that only under specific, generous conditions will the be same problem as what initial parameters produce $y(0) \xrightarrow{} y(\infty)$, the actual desired target solution to the problem. 

% The second limitation of this mapping is that there is redundant information within the description of the model, and the model should be described by the smallest information set possible, to allow the most direct mapping from the output data, to avoid non-unique solutions. In the old formulation, $Pxx$ the input state/parameter covariance matrix changes over the course of the problem. This is redundant, given for this class of models, the final solution is fully described by model equations, initial conditions and parameters. When there is informational redundancy in the model construction, be it with structurally unidentifiable parameters or unnecessarily changing augmented state vector, it leads to a poor mapping from output space to input space, inducing rank deficiency and a solution to a tangential problem. 

\begin{enumerate}
    \item \textbf{Reduction in unnecessary computation of the augmented state mean $\mathbf{\hat{x}}$ and its associated error covariance matrix $\mathbf{P_{\tilde{x}\tilde{x}}}$}: Given that the systems considered here are deterministic, the state estimate $\mathbf{\hat{x}}$ and consequently its error covariance matrix $\mathbf{P_{\tilde{x}\tilde{x}}}$ —do not require iterative updates at every observation point. This is because the initial state $\mathbf{\hat{x}_{0|0}}$, when propagated through the model equations, fully describes the system's trajectory from $\mathbf{y(0) \to y(\infty)}$.

    \item \textbf{Introduction of a Kalman interval $\tau_k$}:
    Given that $\mathbf{\hat x_{0|0}}$, under transformation of the model equations, describes the system $\mathbf{y(0) \xrightarrow{} y(\infty)}$, this process is invariant to the length of discrete time $\tau_k$ which $k$ represents. Therefore, the length of the time for each UKF iteration can be increased, to accommodate more observations. We term this  length of time the \emph{Kalman interval}, denoted $\tau_k$. This holds as the system does not need to be interrupted, to obtain the best estimate of the $\mathbf{\hat x}$ vector, in order to apply a control $\mathbf{u}$, which is the original application of this non-linear filter \cite{julier2004unscented}. 

    \item \textbf{Expansion of the observation vector $\mathbf{y}$ and the innovation covariance matrix $\mathbf{P_{\tilde{y}\tilde{y}}}$}:
    As explained earlier, in a deterministic system without active control, the length of discrete time $\tau_k$ for which $k$ represents can be arbitrarily lengthened, without compromising the application's objective. This extension allows the dimension of the observation vector $\mathbf{y}_{k_\tau}$ to increase accordingly. Now, we define the expanded dimension of the observation vector $\mathbf{y}_{k_\tau}$ to be $a$ such that:
    \begin{equation}
        \label{eq:y_dimension}
        a= \sum_{i=1}^{n} s_i, 
    \end{equation}
    where $n$ still denotes the number of observed outputs (same as in Eq.~(\ref{eq:observation})) and $s_i$ represents the total number of samples for the $i$th observed output in $\tau_k$. In other words, $s_i =\frac{\tau_k}{sf_i}$, where $sf_i$ is the sampling frequency of the $i$th output in Hz.
        Historically, the original UKF method for parameter estimation often suffered from rank deficiency because the number of observations in $\mathbf{y}$ was typically smaller than the number of states and parameters ($\mathbf{ \hat x}$) to be estimated (i.e., $n < L$). By increasing $a$, for this class of system, we overcome this constraint, shifting the limitation from practical data ingestion to the structural identifiability of the system. The necessary condition to overcome rank deficiency is $a \geq L$. In reality, some elements in the observed outputs are correlated, therefore $a$ should be chosen such that $a \gg L$. Note, the number of elements of $\mathbf{P_{\tilde{y}\tilde{y}}}$ $= a^{2}$ and the number of elements in $\mathbf{P_{\tilde{x}\tilde{y}}}$ $= L \times a$.
    
\end{enumerate} 

To the best of our knowledge, our UKF modification is novel. In the sequel, we shall demonstrate its improved ability within the designated setting of estimating parameters and state variables and handling output noise. We shall compare it with the traditional UKF approach, with a suitable application - that of personalising a cardiovascular circulation model. In the process we will, however, emphasise the generalisability of the new approach and its potential in many medical personalisation problems and engineering fields, where accurate estimations of model input parameters are required.

\section{Application to Personalise a 1-Chamber Heart and Circulation Model}

\subsection{The Testbench Reduced-Order Lumped-Parameter Cardiovascular Model}
Our one-chamber (left ventricle) heart and circulation model chosen to evidence our innovative UKF is based on a three-compartmental electrical analogue system by Bjordalsbakke \cite{bjordalsbakke2022parameter}, described by 10 parameters \cite{bjordalsbakke2023monitoring}. See the full description of the model equations in \cite{saxton2023personalised} and in the supplementary material. 

The left ventricle is parameterised by a time-varying elastance function, defined by its minimum ($E_{min}$) and maximum ($E_{max}$) elastance, and fractional timing parameters ($\tau_{es}$, $\tau_{ep}$) which dictate the duration of systole and diastole. The overall heart cycle duration is set by $\tau$ \cite{korakianitis2006numerical,saxton2023personalised}. Flow into and out of the ventricle is controlled by simple diode valves: the mitral valve (with resistance $R_{mv}$) and the aortic valve (with impedance $Z_{ao}$). The other two compartments represent the systemic arterial system, described by an arterial compliance ($C_{sa}$), and the systemic venous system, parametrised by the venous compliance ($C_{sv}$) and the systemic vascular resistance ($R_s$) \cite{saxton2023personalised,bjordalsbakke2022parameter}.

The full parameter vector $\theta$ is:
\begin{equation}
    \label{eq:parameters}
    \mathbf{\theta} = [\tau_{es}, \tau_{ep}, R_{mv}, Z_{ao}, R_s, C_{sa}, C_{sv}, E_{max}, E_{min}, \tau]^T.
\end{equation}

This model has four internal state variables: the pressure in each compartment ($p_{lv}$, $p_{sa}$, $p_{sv}$) and the volume of the left ventricle ($V_{lv}$). A subset of these state variables form the output observable vector $\mathbf{y}$. In our modified UKF implementation, the state vector is augmented to include the parameters as per usual practice. 
The state vector $\mathbf{x}_{\text{state}} \in \mathbb{R}^m$ (where $m=4$) is defined as:
\begin{equation}
    \label{eq:state_vector}
    \mathbf{x}_{\text{state,k}} = [p_{lv}, p_{sa}, p_{sv}, V_{lv}]^\top ,
\end{equation}

and the augmented state vector $\mathbf{x}_k \in \mathbb{R}^{L}$ (where $L = m + p$) is:
\begin{equation}
    \label{eq:augmented_state_vector}
    \mathbf{x}_k = 
    \begin{bmatrix} 
    \mathbf{x}_{\text{state},k} \\ 
    \boldsymbol{\theta}
    \end{bmatrix}.
\end{equation}

The observation vector $\mathbf{y}_k \in \mathbb{R}^a$ represents the set of $a$ available measurement samples (see Eq.~(\ref{eq:y_dimension})) at time $k$, where $a \gg L $. It is related to the state via the observation mapping:
\begin{equation}
    \label{eq:observation_vector}
    \mathbf{y}_k = h(\mathbf{x}_{\text{state},k}) + \mathbf{v}_k
\end{equation}

Here, choosing $a \gg L $ ensures that the sufficient necessary condition for overcoming rank-deficiency is not violated.

\subsection{Synthetic Data Generation}

\subsubsection{Parameter Set Generation}
To create a challenging, representative and diverse dataset, to robustly test the modified UKF method, 50 target parameter sets  ($\mathbf{\theta_{\text{target}}}$) were generated by sampling from a truncated normal (Gaussian) distribution. 

First, upper ($ub$) and lower ($lb$) bounds for each parameter were defined as $\pm 60\%$ of the expectation value of the Gaussian distribution from which the target parameter sets were sampled from:
\begin{align}
    lb &= \theta_{\text{normal}} (1 - 0.6) = 0.4  \theta_{\text{normal}} \label{eq:param_lb} \\
    ub &= \theta_{\text{normal}}  (1 + 0.6) = 1.6 \theta_{\text{normal}} \label{eq:param_ub}
\end{align}
This was the most stringent test that could be applied: when going above 60\%, the solver failed to reliably produce working simulations as the parameter 
combinations became non-physiological. 
A mean ($\mu$) and standard deviation ($\sigma$) were then calculated from these bounds. The mean was set to the midpoint of the range (i.e., $\theta_{\text{normal}}$), and the standard deviation was set such that the bounds represent $\pm 3\sigma$ (covering approximately $99.7\%$ of the distribution):
\begin{align}
    \mu &= \frac{ub + lb}{2} = \theta_{\text{normal}} \label{eq:param_mean} \\
    \sigma &= \frac{ub - lb}{6} = \frac{1.2  \theta_{\text{normal}}}{6} = 0.2  \theta_{\text{normal}} \label{eq:param_std}
\end{align}

Each target parameter was then drawn from this distribution:
\begin{equation}
    \label{eq:param_generation}
    \theta_{\text{target}} \sim \mathcal{N}(\mu, \sigma^2) = \mathcal{N}(\theta_{\text{normal}}, (0.2  \theta_{\text{normal}})^2)
\end{equation}

Any parameter sets for which the ODE solver was unable to produce stable solutions, or for which the results were non-physiological, were discarded. This process was repeated until 50 viable parameter target sets were obtained. This was to ensure only input parameters leading to clinically viable ODE solutions of the model would to be estimated by the UKF later. 

\subsubsection{Plausibility Check for Parameter Sets}
To ensure that the sampled parameter sets resulted in physiologically plausible cardiovascular states, the ODE solver's output for each set was checked against a set of relaxed clinical criteria (the purpose of this was mainly to stop the solver becoming unphysiological and failing). A parameter set was only considered viable if all resulting key physiological variables from the steady-state simulation fell within the predefined ranges, as summarised in Table \ref{tab:plausibility_limits}.

\begin{table}[ht]
    \centering
    \caption{Relaxed physiological limits used to determine parameter set plausibility.}
    \label{tab:plausibility_limits}
    \begin{tabular}{lccc}
        \toprule
        \textbf{Variable} & \textbf{Lower Bound} & \textbf{Upper Bound} & \textbf{Unit} \\
        \midrule
        Systolic BP (SBP, $P_{sa,sys}$) & 60 & 250 & mmHg \\
        Diastolic BP (DBP, $P_{sa,dia}$) & 30 & 150 & mmHg \\
        LV End-Diastolic Pressure ($P_{lv,dia}$) & N/A & 40 & mmHg \\
        Stroke Volume (SV) & 20 & 180 & mL \\
        Ejection Fraction (EF) & 0.15 & 0.85 & Unitless \\
        Cardiac Output (CO) & 2.0 & 12.0 & L/min \\
        \bottomrule
    \end{tabular}
\end{table}

These limits represent a significantly wider range than healthy human physiology, deliberately including various pathological states to challenge the UKF and ensure the resulting ensemble was diverse. For example, the criteria included states ranging from severe hypotension ($P_{sa,sys} \ge 60$ mmHg) to hypertensive crisis ($P_{sa,sys} \le 250$ mmHg).

% \subsubsection{Abbreviations for Clinical Variables}
% The following abbreviations are used throughout the text when discussing the clinical state of the generated data.

% \begin{longtable}{lll}
%     \caption{Abbreviations for key clinical and physiological variables.}
%     \label{tab:abbreviations} \\
%     \toprule
%     \textbf{Abbreviation} & \textbf{Description} & \textbf{Unit} \\
%     \midrule
%     \endfirsthead
%     \toprule
%     \textbf{Abbreviation} & \textbf{Description} & \textbf{Unit} \\
%     \midrule
%     \endhead
%     SBP & Systolic Blood Pressure & mmHg \\
%     DBP & Diastolic Blood Pressure & mmHg \\
%     EF & Ejection Fraction & \% \\
%     HR & Heart Rate & bpm \\
%     % LVEDP & Left Ventricular End-Diastolic Pressure & mmHg \\
%     % HFpEF & Heart Failure with Preserved Ejection Fraction & N/A \\
%     \bottomrule
% \end{longtable}

\subsubsection{Ensemble Characteristics}
\label{sec:ensemble}
The resulting 50 parameter sets produced a wide range of clinically relevant patho-physiologies \cite{david2005taylor,nichols2022mcdonald}. A summary of the key conditions present in the ensemble is shown in Table \ref{tab:pathophysiology_summary}.

\begin{table}[ht]
    \centering
    \caption{Distribution of patho-physiological conditions across the 50 target parameter sets. Note that some parameter sets may exhibit multiple conditions, likewise some have no patho-physiology.}
    \label{tab:pathophysiology_summary}
    \begin{tabular}{llc}
        \toprule
        \textbf{Pathophysiological Condition} & \textbf{Clinical Criteria} & \textbf{Count (\%)} \\
        \midrule
        Hypotension & (SBP $<$ 90 or DBP $<$ 60 mmHg) & 7 (14.00\%) \\
        Hypertension Stage 2 & (SBP $\geq$ 140 or DBP $\geq$ 90 mmHg) & 13 (26.00\%) \\
        Hypertensive Crisis & (SBP $>$ 200 or DBP $>$ 120 mmHg) & 1 (2.00\%) \\
        % Bradycardia & (HR $<$ 60 bpm) & 0 (0.00\%) \\
        % Tachycardia & (HR $>$ 100 bpm) & 0 (0.00\%) \\
        Heart Failure w/ Reduced EF & (EF $<$ 40\%) & 9 (18.00\%) \\
        % Indicators of HFpEF & (EF $\geq$ 50\% \& LVEDP $>$ 15 mmHg) & 0 (0.00\%) \\
        Wide Pulse Pressure & ($>$ 60 mmHg) & 16 (32.00\%) \\
        \bottomrule
    \end{tabular}
\end{table}

\subsubsection{Output Signal Generation}
For each of the 50 target parameter sets, the model was run to a steady state first. The outputs (observables) were then sampled with the same frequency as the non-adaptive ODE solver. To replicate realistic clinical data, zero-mean Gaussian multiplicative noise ($\epsilon$) was applied to the true signal ($y_{\text{true}}$):
\begin{equation}
\label{eq:noise}
    \mathbf{y}_{\text{obs}} = (1 + \epsilon) \mathbf{y}_{\text{true}}, \quad \text{where} \quad \epsilon \sim \mathcal{N}(0, \sigma_{\text{noise}}^2).
\end{equation}

The noise standard deviation, $\sigma_{\text{noise}}$, was set to $1\%$, $5\%$, and $10\%$ for different simulation runs. A smoothing function was then applied to the noisy data ($y_{\text{obs}}$), in line with typical clinical data ingestion techniques:
\begin{equation}
    \label{eq:smoothing_function}
    y_{\text{obs},i} = \frac{1}{u_i - l_i + 1}
    \sum_{j=l_i}^{u_i} y_{\text{obs},j}
\end{equation}

where the lower and upper bounds are defined as:
\begin{align*}
    l_i &= \max(1, i - w/2) \\
    u_i &= \min(N, i + w/2)
\end{align*}
and $w$ represents the window size which is set to 5 here.

\subsubsection{Initial Parameter Guess and Robust Methodological Construction}

The initial guess of each parameter was chosen to be the mean of the underlying distribution from which the parameter sets were randomly sampled from. This is a very robust choice for clinical application, as it infers little information to provide to the algorithm about the target parameter sample, which is typically completely unknown. 

Similarly, the initial error covariance matrix, $\mathbf{P_{\tilde{x}\tilde{x}}}_{0|0}$, was not specifically tuned, which has been a constraint of the traditional UKF implementations. Researchers have noted that "\textit{a priori knowledge, the initial covariance matrix, the noise covariance matrices are most important to the performance and stability of the UKF}" \cite{majeed2013aerodynamic} . In this implementation, the initial covariance matrix stayed the same for every estimation attempt which implies no prior knowledge of the parameter targets. This methodological construction is much more robust, and is designed to test the method's applicability to real-world applications where the target parameters (ground truths) and noise variances (for both the process noise and the observation noise) are unknown to the practitioner. 

\subsection{Application of UKF to the Circulation Model}

The first step is to create the augmented state vector, consisting of 4 internal states: $p_{lv}$, $p_{la}$, $p_{sv}$ and $v_{lv}$ (see Eq.~(\ref{eq:state_vector})) and the 10 parameters (see Eq.~(\ref{eq:parameters})). This 14-element augmented state vector has an associated initial covariance matrix $\mathbf{P_{0|0}}$ describing the probability distribution of each element, and how these probability distributions are conditioned on each other \cite{wan2000unscented, saxton2023personalised}. 

The next step is to choose $2L+1$ sigma points (here $L=14$). These sigma points are chosen based on the covariance matrix, with points chosen further from the mean if the covariance is higher; the distribution of the other 13 parameters and states are altered to reflect how the probability distributions are conditioned \cite{wan2000unscented, saxton2023personalised}.

Next, importantly, the Kalman interval $\tau_{k}$ was chosen to be one heart cycle $\tau$, which was typically $1s$ though this parameter $\tau$, was left to vary during estimation. The system was observed at the interval of the ODE solver $\delta t= 0.001$, which is typical of many attempts at this problem \cite{saxton2023personalised}. Therefore, the length of the vector $\mathbf{y}$ would equal $ n \times \tau_{k}/\delta t$ where $n$ is the number of observed states. The size of $n$ varied to test the identifiability of the model. 

The 29 sigma points are propagated through the model at the end of interval $\tau_{k}$ (which we named as `the Kalman interval' in section~\ref{sec:modified_approach}). This allows the model to form predictions around the distribution of the outputs $\mathbf{y}$ given the distribution of the state error $\mathbf{P_{\tilde{x}\tilde{x}}}_{k|k-1}$, which are captured discretely through sigma points. This allows summary statistics $\mathbf{P_{\tilde{x}\tilde{y}}}_{k|k-1}$ and  $\mathbf{P_{\tilde{y}\tilde{y}}}_{k|k-1}$ to be calculated, which contain the information in order to update the posterior distribution of the augmented state error $\mathbf{P_{\tilde{x}\tilde{x}}}_{k|k}$. This is performed via the Kalman gain matrix $\mathbf{K}$, which is calculated optimally and used to project information from the output domain back into the input domain, to form better estimates of model states and parameters.

\section{Results and Discussions}
The modified UKF algorithm was run for 100 cycles, requiring modest computational resource- it took only 40 seconds to run the 100 cycles of the algorithm on a research laptop, under the assumption that one time series of the output observation is available (i.e., $n=1$). The computational time increases with the square of the number of output observations (i.e., $O(a^2)$); accordingly, 4 outputs takes approximately 16 times longer (around 10 minutes). Again, this is a modest computational overhead, given that this is a modest research laptop 
(specifications: Processor 13th Gen Intel(R) Core(TM) i7-13650HX (2.60 GHz), RAM 16.0 GB). Note, the computations are performed in the Julia programming language.

\subsection{Generalisability for Parameter Estimation}
In this section, we demonstrate the generalisability of the new method for parameter estimation. As explained in Section~\ref{sec:ensemble}, there are 50 unknown target parameter sets, some of which describe extreme pathophysiology. Despite the diversity of the parameter sets, in our implementation of the modified UKF method, specific empirical tuning of the initial error covariance matrix $\mathbf{P_{\tilde{x}\tilde{x}_{0|0}}}$ or the initial state estimate $\mathbf{\hat{x}_{0|0}}$ is not required to achieve a high degree of parameter estimation convergence. The new method is largely robust to the lack of a priori knowledge regarding the target parameter set.

\subsubsection{Identifiability Heatmaps with 1\% Noise (98\%, 95\% and 90\% Accuracy)}
\begin{figure}[htbp]
    \centering
    % Ensure this filename matches your uploaded file
    \includegraphics[width=\textwidth]{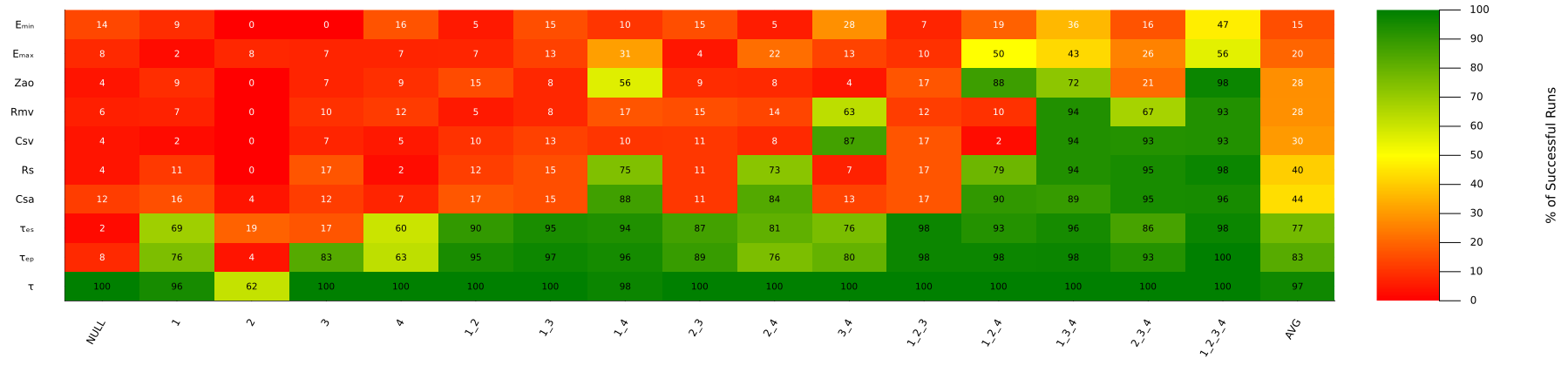}
    \caption{Identifiability heatmap for $\geq$ 98\% accuracy at 1\% noise level ($\sigma_{\text{noise}} = 0.01$). Each number in the heatmap represents the percentage of parameter sets which achieved a minimum of $98\%$ of accuracy at 1\% noise, for each parameter, from the available observation combinations.}
    \label{fig:heatmap_noise_98acc_0.01noise}
\end{figure}

\begin{figure}[htbp]
    \centering
    % Ensure this filename matches your uploaded file
    \includegraphics[width=\textwidth]{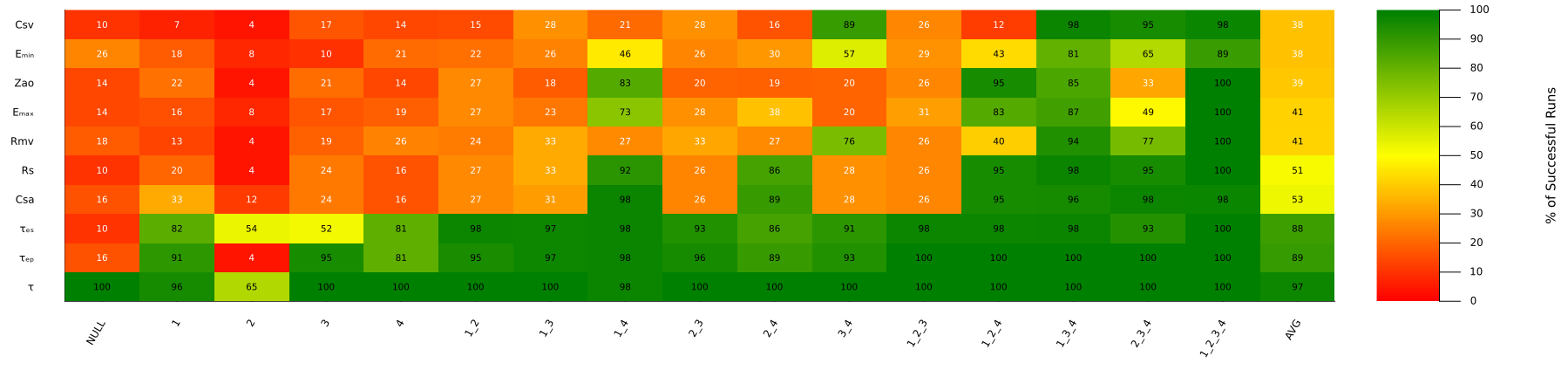}
    \caption{Identifiability heatmap for $\geq$ 95\% accuracy at 1\% noise level ($\sigma_{\text{noise}} = 0.01$). Each number in the heatmap represents the percentage of parameter sets which achieved a minimum of $95\%$ of accuracy at 1\% noise, for each parameter, from the available observation combinations.}
    \label{fig:heatmap_noise_95acc_0.01noise}
\end{figure}

\begin{figure}[htbp]
    \centering
    % Ensure this filename matches your uploaded file
    \includegraphics[width=\textwidth]{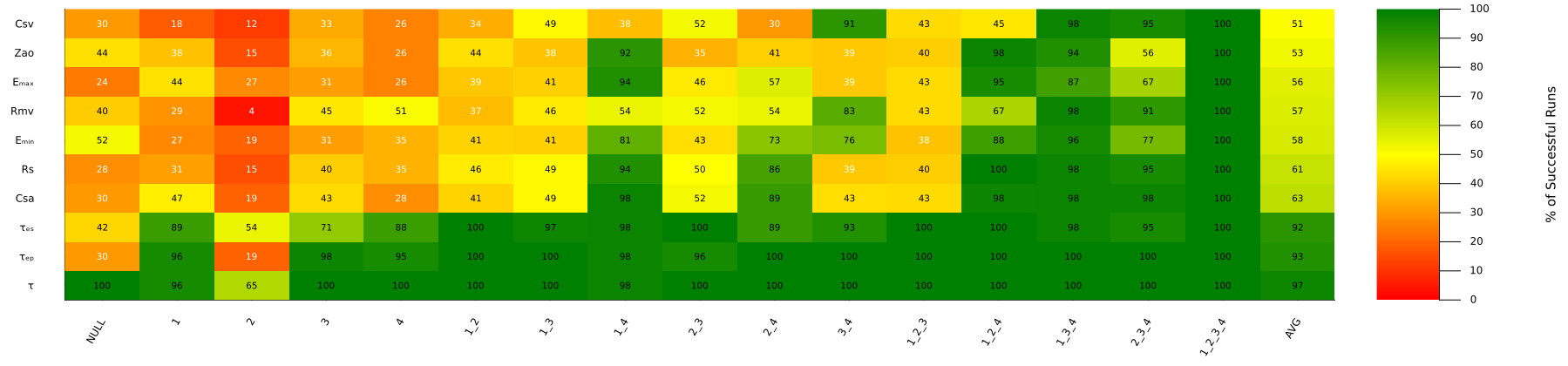}
    \caption{Identifiability heatmap for $\geq$ 90\% accuracy at 1\% noise level ($\sigma_{\text{noise}} = 0.01$). Each number in the heatmap represents the percentage of parameter sets which achieved a minimum of $90\%$ of accuracy at 1\% noise, for each parameter, from the available observation combinations.}
    \label{fig:heatmap_noise_90acc_0.01noise}
\end{figure}

The following three figures, Figures~\ref{fig:heatmap_noise_98acc_0.01noise}--\ref{fig:heatmap_noise_90acc_0.01noise}, show the percentage of successful runs that achieved an accuracy of $\geq 98\%$ , $\geq 95\%$ and $\geq 90\%$, respectively, for each parameter (vertical-axis) across all tested observation index combinations (horizontal-axis), all of which subject to 1\% noise (i.e., $\sigma_{\text{noise}} = 1\%$ in Eq.~(\ref{eq:noise})). Here, the pressures $p_{lv}$, $p_{sa}$, $p_{sv}$ and the volume of the left ventricle $V_{lv}$ refer to indices 1, 2, 3 and 4 respectively. The null results quantify the number of simulations in which the naive initial guess fell within the specified accuracy threshold of the target parameter.

The full observation set, $p_{lv}$, $p_{sa}$, $p_{sv}$ and $V_{lv}$ $(1,2,3,4)$, converged to at least 98\% accuracy for over 90\% of model runs, for the parameters $\tau_{es}, \tau_{ep}, R_{mv}, Z_{ao}, R_s, C_{sa}, C_{sv}$ and $\tau$. This shows a remarkable level of recovery of these parameters, considering the initial guesses (first column in Figure~\ref{fig:heatmap_noise_98acc_0.01noise}) are within 98\% accuracy on the true solution, on average, under 10\% of the time (apart from $\tau$). The modified UKF implementation recovered $E_{max}$, $E_{min}$ to 98\% accuracy in $\approx 50\%$ of simulation runs. 

\clearpage

When the accuracy criterion is loosened to 95\%, however, the full observation set $p_{lv}$, $p_{sa}$, $p_{sv}$ and $V_{lv}$ $(1,2,3,4)$ shows almost complete recovery in every simulation, as shown in Figure~\ref{fig:heatmap_noise_95acc_0.01noise}. 
Interestingly, the observation set $p_{lv}$ and $V_{lv}$, $(1,4)$, recovered the parameters $\tau_{es}, \tau_{ep},  R_s, C_{sa}$ and $ \tau $ to 95\% accuracy over 90\% of the time and $Z_{ao}$ and $E_{max}$ over 70\% of the time. It partially recovered $E_{min}$, and struggled to recover $C_{sv}, R_{mv}$ much above the baseline null guess. This result has significant implications - only two observations from the left ventricle $(1,4)$ are provided to the modified UKF, yet it is able to estimate parameters $Z_{ao}, R_s, C_{sa}$, which are not contained within this compartment. These results are further demonstrated in Figure~\ref {fig:heatmap_noise_90acc_0.01noise}, where the accuracy criterion is set as 90\%.

Figure~ \ref{fig:heatmap_noise_90acc_0.01noise} also illustrates the importance of combining a pressure observation with the volume observation to achieve parameter estimation reliably with the modified UKF. With single or combined pressure observations alone, even the modified UKF cannot recover the parameters reliably above the null guess, other than the timing parameters. 

% \FloatBarrier 
% \subsection{Noise Comparison for 4-Observable Input}
% This figure provides a direct comparison of parameter identifiability across the three noise levels (1\%, 5\%, 10\%) and the null guess. This analysis shows the percentage of parameter sets which converged to 98\% accuracy, broken down by parameter. This is for the full 4-observable input set ($p_{lv}$, $p_{sa}$, $p_{sv}$, $V_{lv}$).

% \begin{figure}[htbp]
%     \centering
%     % Ensure this filename matches your uploaded file
%     \includegraphics[width=0.9\textwidth]{accuracy_report_idxs_1_2_3_4_98.0pct.png}
%     \caption{Identifiability comparison for the 4-observables ($p_{lv}$, $p_{sa}$, $p_{sv}$) and ($V_{lv}$)  at 98\% accuracy.}
%     \label{fig:barchart_1234}
% \end{figure}

% \FloatBarrier 

\subsection{Comparisons with the Original UKF Implementation}

 In order to compare the modified UKF methodology with the original UKF approach for parameter estimation, the latter was implemented for 100 cycle runs on the same 10-parameter, single-chamber heart circulation model. The innovation covariance matrix is small and set at every observation point, $\delta t = 0.001\text{ s}$. The initial error covariance matrix $\mathbf{P}_{0|0}$ was set identically to that used in the modified method and was not tuned. The parameter sets and initial estimates were all configured to be the same as those employed to test the modified method, with the measurement noise set at $5\%$. This study constitutes arguably the most rigorous assessment of the performance of the original UKF methodologies reported thus far. A diverse target parameter set, comprising 50 samples, was utilised, featuring up to a $60\%$ discrepancy from the initial estimate, compounded by a $5\%$ magnitude of measurement noise, and critically, without any specialised or time-intensive calibration of initial conditions, such as the initial covariance. The original approach demonstrated a comprehensive breakdown when subjected to this level of stress. We propose that this vulnerability is the fundamental reason why researchers, to our knowledge, have been unable to achieve results demonstrating high convergence fidelity on large, diverse datasets characterised by a high dimensionality of free parameters and significant noise perturbation. The established consensus within the literature is that these methods lack robust invariance concerning factors such as noise level, initial conditions (including starting parameter estimates and covariance specifications), and the size of the free parameter set.

\subsubsection{Convergence Plots - Original UKF Method vs New UKF Method }
The following figure, Figure~ \ref{fig:paramter_convergence}, shows the convergence of the parameter set with 5\% noise applied for each observation, for the full observable set ($p_{lv}$, $p_{sa}$, $p_{sv}$ and $V_{lv}$). Parameter estimation results, deriving from the original and the new UKF methods are compared. The initial guess was set to be the same as in the case of the generalised ensemble of results - it is the expectation value of the Gaussian distribution from which the samples were drawn from. We can see from the results that the original UKF implementation is unstable and struggles to converge on parameter estimates, under the severe stress of the current conditions. 

\begin{figure}[htbp]
    \centering
    % Ensure this filename matches your uploaded file
    \includegraphics[width=\textwidth] {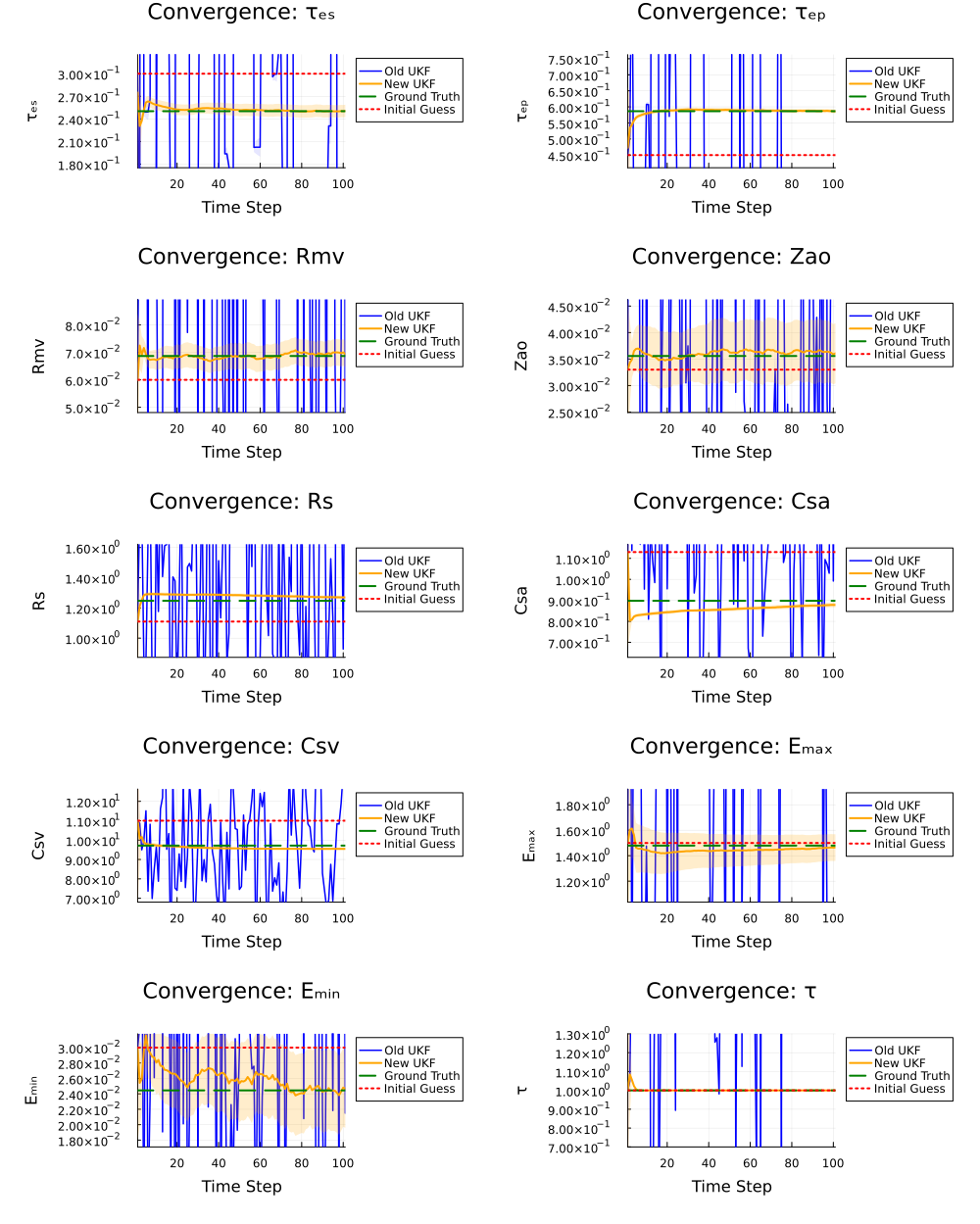}
    \caption{The figure shows the comparison of the original (old) and modified (new) UKF implementations on the convergence of the parameter sets over the duration of 100 cycles at 5\% observation noise for the full observable set ($p_{lv}$, $p_{sa}$, $p_{sv}$ and $V_{lv}$). The shaded orange and blue regions represent the variance of the parameter estimate. For the new method, with parameters 
    $R_s, C_{sa}$ and $C_{sv}$, there has been covariance collapse, such that the filter has become overconfident on its estimates. To reiterate $\tau$ was a free parameter during tuning, however all target sets had $\tau$ set to 1.}
    \label{fig:paramter_convergence}
\end{figure}

\clearpage

\subsection{Robustness for Parameter Estimation}
In this section, we demonstrate the new method's robustness under noise stress. The result holds well when the noise level is increased from 1\% to 5\% and still holds fairly well at 10\%. This is important for clinical translation, as data gathering devices are typically very noisy: from $2\% \xrightarrow{} 10\%$  \cite{grothues2002comparison,bellenger2000comparison,magder2015central,figg2009error,gardner1981direct}. 
% For a full set of results, see Appendix \ref{Appendix:A}.

\begin{figure}[htbp]
    \centering
    % Ensure this filename matches your uploaded file
    \includegraphics[width=0.9\textwidth]{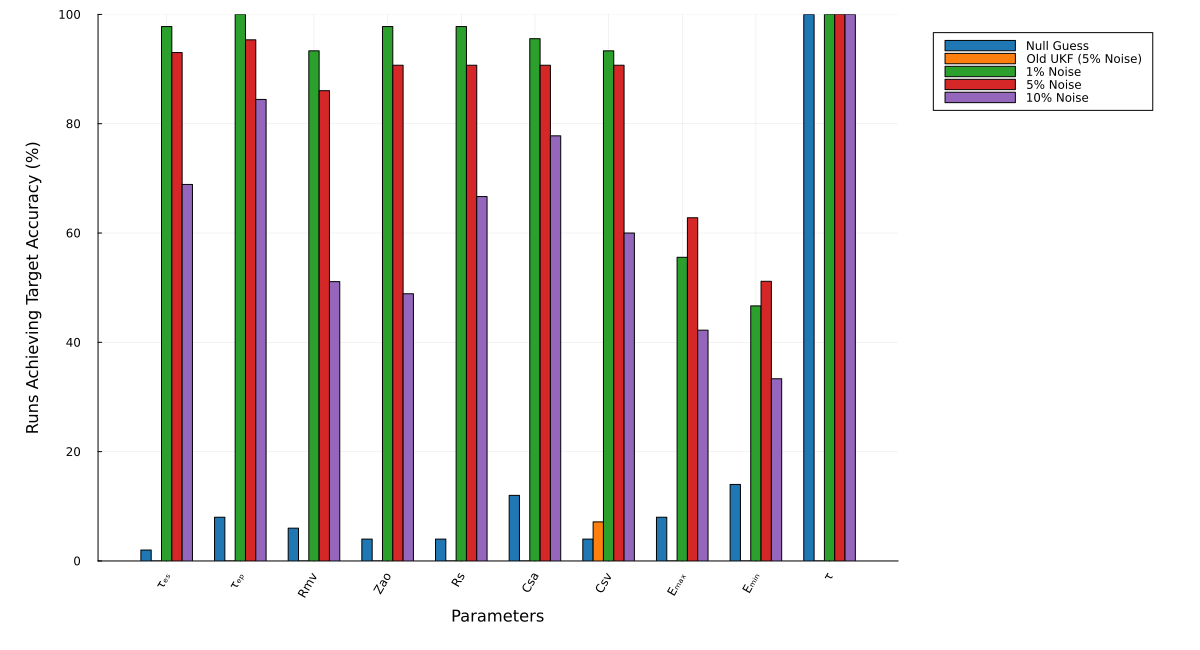}
    \caption{Identifiability comparison for the 4-observables ($p_{lv}$, $p_{sa}$, $p_{sv}$) and ($V_{lv}$)  at $\geq98\%$ accuracy, with varying levels of noise, against the original method with 5\% noise. To reiterate $\tau$ was a free parameter during tuning, however all target sets had $\tau$ set to 1.}
    \label{fig:barchart_1234_old_method}
\end{figure}

%\subsubsection{Noise Comparison for 4-Observable ($p_{lv}$, $p_{sa}$, $p_{sv}$ and $V_{lv}$) output set, original and new method}
Figure~\ref{fig:barchart_1234_old_method} provides a comparative analysis of parameter identifiability between the modified UKF across three noise levels (1\%, 5\%, and 10\%) and the original UKF at a 5\% noise threshold. The null initial guess is included as a baseline reference point to contextualise the convergence performance. Identifiability is defined as the percentage of the parameter ensemble that achieved a convergence accuracy of $\geq 98\%$, utilising the complete four-observable set ($P_{lv}, P_{sa}, P_{sv}, V_{lv}$). Notably, the original UKF is represented by a single bar at 5\% noise; the absence of additional data points for this method is a result of numerical divergence and a loss of filter stability under the prescribed noise conditions, rather than an error in the data representation. Detailed results in the form of heatmaps are shown in the supplementary material.

% \FloatBarrier 
%\clearpage

\subsection{State Estimation} 
Beyond parameter identification, the estimation of unobserved internal states is also a clinically significant application of the modified UKF. In clinical practice, measuring every haemodynamic variable is often unfeasible due to the associated risks to patients, invasiveness, and the costs of additional instrumentation and personnel. Model-based state estimation allows for the ``blind'' reconstruction of these latent variables using a minimum set of measurement instruments. While conventional techniques such as the EKF often face stability issues when dealing with highly nonlinear discontinuities, such as valve dynamics, the UKF provides a robust framework for inferring unmeasured states of highly nonlinear dynamics.

\begin{figure}[!ht]
    \centering
    \captionsetup[subfigure]{justification=centering} % Centers subcaptions
    
    % Use the \foreach loop to iterate from 1 to 9
    \foreach \n in {1,...,12}{
        \begin{subfigure}{0.32\textwidth} % Set width to 0.32\textwidth (approx 1/3 page width)
            \centering
            % Include the plot. The width is relative to the subfigure width (0.32\textwidth)
            \includegraphics[width=\textwidth]{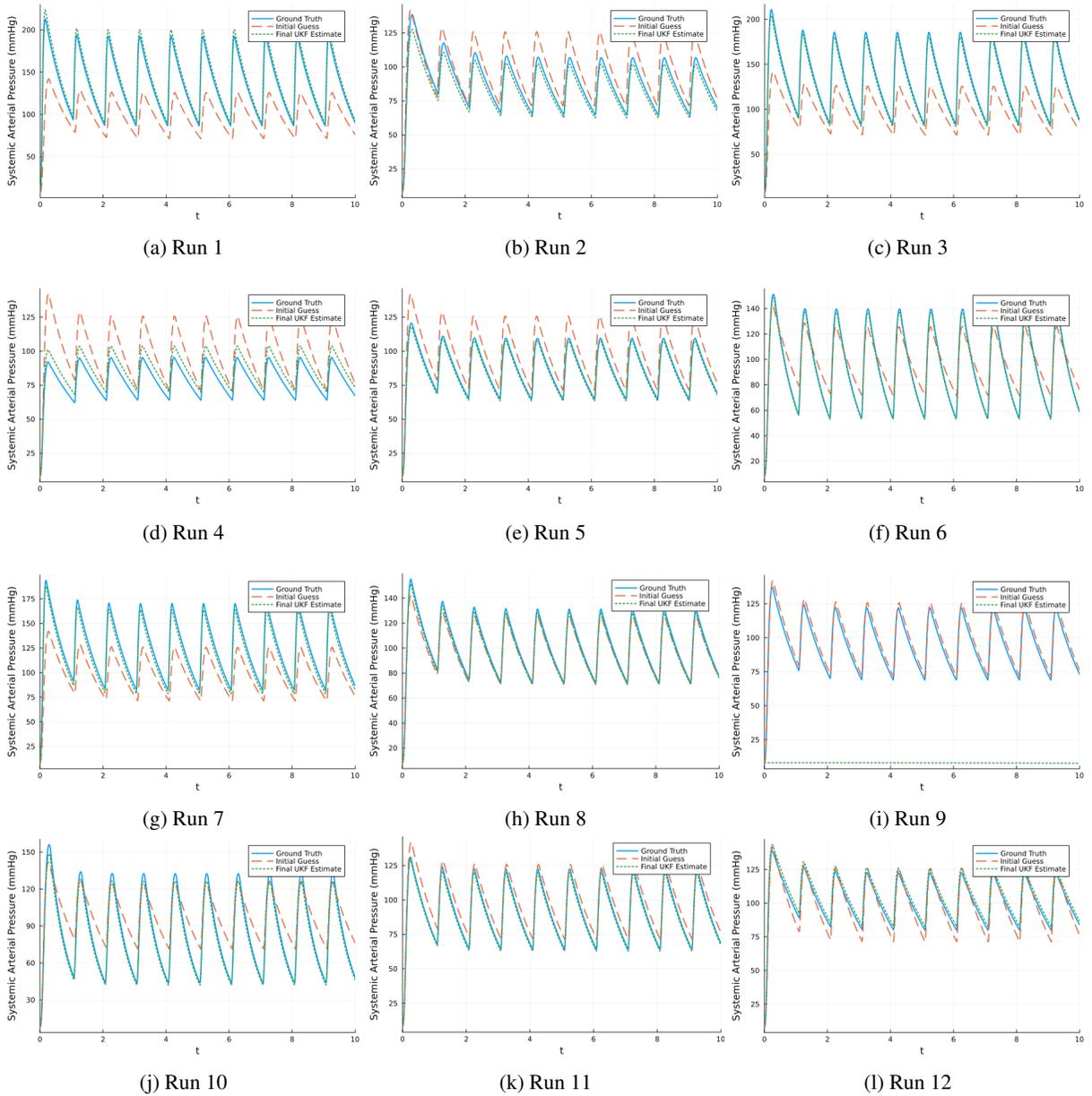}
            \caption{Run \n}
            \label{fig:run_\n}
        \end{subfigure}%
        % -------------------------------------------------------------
        % Manual Row Breaks: Start a new row after the 3rd and 6th plots.
        \ifnum\n=3 \par\vspace{1em}\fi
        \ifnum\n=6 \par\vspace{1em}\fi
        % -------------------------------------------------------------
    }
    
    \caption{Blind State estimation comparison for 12 randomly selected parameter estimation runs with 1\% noise. The graphs are showing the estimates for systemic arterial pressure $p_{sa}$, when the modified UKF has only been shown $p_{lv}$ and $V_{lv}$ which is output set (1,4) on the heatmap Figure~\ref{fig:heatmap_noise_98acc_0.01noise}.}
    \label{fig:selected_runs_comparison}
\end{figure}

To illustrate this point, we evaluated our filter's ability to reconstruct the systemic arterial pressure ($P_{sa}$) without direct measurement. Using only observations of the left ventricular pressure ($P_{lv}$) and volume ($V_{lv}$), the modified UKF successfully estimated the unobserved arterial pressure waveform (see Figure \ref{fig:selected_runs_comparison}). This demonstrates the potential of our method to infer downstream haemodynamic conditions from limited upstream data. 

To rigorously assess the filter's performance in reconstructing unobserved variables, the modified UKF was deliberately deprived of any measurement information pertaining to the systemic arterial system ($P_{sa}$), relying solely on the observed left ventricular variables ($P_{lv}$ and $V_{lv}$). Despite this significant constraint and the challenging parameter space which resulted in a reduced convergence rate for the full parameter vector (as quantified by scores below 100\% in the parameter accuracy heatmap, Figure~\ref{fig:heatmap_noise_98acc_0.01noise} (output 1,4)). The UKF consistently captured the systemic arterial pressure waveform with high fidelity. Across the tested runs, the estimated $P_{sa}$ trajectory accurately reproduced the true arterial pressure, failing to provide a viable estimate in two instances: Run 4 (Figure~\ref{fig:run_4}), where slight divergence from the true trajectory was observed, and Run 9 (Figure~\ref{fig:run_9}), where the underlying ODE solution failed to produce a stable output.

\clearpage

\subsection{Discussions}
Overall, these results demonstrate substantial improvements and provide strong evidence for the efficacy of the modified UKF in personalised modelling. 
Our approach is shown to be, in practical terms, fully identifiable, and given the properties of the model to which it is applied here, generalisable. This is true, even with large amounts of noise and varied target data representative of both pathology and patho-physiology. 
A notable property of our method is its ability to reconstruct underlying biological mechanics, allowing for the determination of the information contained within a certain set of observations. At a 1\% noise level, where information is least corrupted, operating with only two model outputs $V_{lv}$ and $p_{lv}$ allows for the successful recovery of left ventricle timing parameters. Interestingly, one can also accurately back-out systemic arterial compliance $C_{sa}$ and systemic resistance $R_s$, while achieving more modest accuracy for minimum elastance $E_{min}$ and aortic valve impedance $Z_{ao}$. The only testbench model input parameters which our UKF recovered with limited success were venous compliance $C_{sv}$ and mitral valve resistance $R_{mv}$. The results are less promising when only one observation or two pressure observations are available to the model and the modified UKF algorithm. This is because it is not mathematically possible to derive parameters such as resistance or compliance without a flow (derivative of volume) and pressure measurement, given the model equations. 

Another surprising result is that there is barely any degradation in the modified UKF's performance when using 5\% noise compared to 1\%. Significantly, at 10\% noise, the use of the full testbench model's set of outputs allowed us to recover all of the input parameters to 98\% accuracy, in over half the tests. 
In addition, promising results emerge from the three observable set $\{V_{lv}, p_{sa}, p_{sv} \}$ at a noise level of 1\% and 95\% convergence, which data show that almost all parameters with the exception of aortic impedance converged to this practical, reasonable level of accuracy, without the support of invasive left ventricle pressure measurements. This could potentially identify an important diagnostic observation set for clinicians, given that venous and arterial pressure are fairly straightforward measurements. 
This algorithmic method will also offer a framework for questioning the structural identifiability of models.

\section{Conclusion} 
Our modified method recovered a diverse target dataset of 50 samples with a ``normal'' distribution and a standard deviation of 20\%. Each sample contains all 10 parameters and the samples represent a range from healthy physiology to extreme patho-physiology, with minimum / maximum values capped at 60\% away from the mean average. This value was chosen to make the recovery as difficult as possible. The modified UKF algorithm is robust in recovering almost all parameters to over 98\% accuracy, over 90\% of the time, even with a moderate 5\% multiplicative noise for the complete output set of left ventricle pressure $p_{lv}$, systemic arterial pressure $p_{sa}$, systemic venous pressure $p_{sv}$ and left ventricle volume $V_{lv}$. This result still holds fairly well for 10\% noise, where all parameters converged to at least 95\% accuracy on more than 60\% of simulations. Perhaps even more impressively, the above results are obtained from a simple algorithmic implementation, for example, being limited to only 100 cycles. Another surprising and interesting finding is that most parameters are recovered to this high level of accuracy, even with just two observations $p_{lv}$ and $V_{lv}$. We compare this to the original implementation of this algorithm for parameter estimation and believe we have made a significant improvement. The modified method excels in recovering patho-physiology, and contributes to practical solutions of the personalisation problem for patient-specific healthcare design \cite{pant2017inverse,saxton2023personalised,saxton2024assessing,saxton2025impact}.

The findings presented here offer a significant advancement in cardiovascular personalisation. This research also provides methodological insights relevant to the wider context of parameter estimation. It overcomes three critical issues in parameter estimation with the UKF: rank deficiency of the covariance matrices, robustness under the ignorance of initial conditions (tuning of covariance), and a miss attribution of the target problem \cite{canuto2020ensemble}. 
The generalisability and robustness of the parameter estimation under diverse target patho-physiology and high noise stress is encouraging for clinical applications. 

The method introduced in this paper represents a considerable development of the family of gradient-free parameter estimation methods, and utilises the information within the data to a much greater extent than previous applications. The $\mathbf{P_{\tilde{y}\tilde{y}}}$ matrix uses correlations between observations of $\mathbf{y}$ at different points in time, allowing the shape of the output data of the model to be effective in the estimation of the parameters.

The main limitation in applying this method clinically is that the model output data used in our example in this paper are a construction of the model itself, even though noise is considered and added to the synthetic data. Further work will be needed to appropriately handle distortions or transformations that come from the measurement of observables or a model that cannot fully reproduce the underlying observed physiology. This will modify the interpretation of parameters, and also in turn affect the estimation of those parameters. When using lumped parameter models, researchers and clinicians will need to understand the translation of these complex model parameters onto human physiology. For example, mitral valve resistance in this model could also represent many defects with the atria, this point can not be stressed enough for these low complexity lumped models.

\section*{Research Funding Information}
A.T. is funded by the University of Sheffield PhD scholarship scheme.
X.X. and I.H. are funded by EPSRC Health Technologies Connectivity Awards (UKRI830). 

\bibliographystyle{plos2015}  
\bibliography{references}  

\end{document}